\documentclass[oribibl]{llncs}

\usepackage{amsmath}
\usepackage{amsfonts}

\usepackage{epsfig}

\usepackage{index}
\newindex{nam}{ndx}{nnd}{Name Index}

\newcommand{\mb}[1]{\mbox{\boldmath $#1$}}
\newcommand{\mat}[1]{\mathrm{#1}}

\newcommand{\Idmat}{\mb{1}}
\renewcommand{\eqref}[1]{Eq.~(\ref{#1})}

\begin{document}

\title{Molecular Evolution in Time-Dependent Environments}
\titlerunning{Time-dependent replication landscape}  
%
\author{Claus O.~Wilke \and Christopher Ronnewinkel \and Thomas Martinetz}

\institute{Institut f\"ur Neuroinformatik, Ruhr-Universit\"at Bochum\\
  D-44780 Bochum, Germany\\
email: \texttt{wilke,ronne,martinetz@neuroinformatik.ruhr-uni-bochum.de},\\
WWW home page: \texttt{http://www.neuroinformatik.ruhr-uni-bochum.de}
}

\maketitle              

\begin{abstract}
The quasispecies theory is studied for dynamic replication
landscapes. A meaningful asymptotic quasispecies is defined for
periodic time dependencies. The quasispecies' composition is constantly
changing over the oscillation period. The error threshold moves
towards the position of the time averaged landscape for high
oscillation frequencies and follows the landscape closely for low
oscillation frequencies.
\end{abstract}

The quasispecies theory, put forward by Eigen in 1971~\cite{Eigen71},
and subsequently studied by Eigen, Schuster, McCaskill and
coworkers~\cite{EigenSchuster79,Eigenetal88,Eigenetal89}, is nowadays
one of the classical theories of self-replicating entities. Its
prediction of an error threshold, above which the self-replication
ceases to produce useful offspring, has important implications for the
origin of life. The error threshold effectively limits the amount of
information the entities can carry, thus placing an upper bound on the
complexity self-reproducing information carriers can achieve without
sophisticated error correction mechanisms.

Although completely static environments are unrealistic in any case
apart from experiments in perfectly controlled flow reactors, the
quasispecies theory has so far been considered mainly in static
replication landscapes. Nevertheless, even under fixed environmental
conditions can the replication rates of RNA molecules, for example,
change because of changing concentrations of template and
replica~\cite{Biebricheretal83}. Jones~\cite{Jones79a,Jones79b} has studied underlying
time-dependencies which are identical for all sequences. 
Contrasting to that, we want to focus on replication landscapes with
individual time-dependency for each sequence. One of the reasons for
the neglect of 
individually changing replication coefficients in earlier work is
probably the fact that for 
arbitrary temporal changes an asymptotic quasispecies cannot be
defined. However, a meaningful definition is at hand for time-periodic
replication landscapes, as we are going to show below.

We start from the discretized form of Eigen's evolution
equation~\cite{Demetriusetal85}, linearized with the appropriate
transformation~\cite{ThompsonMcBride74,JonesEnnsRangnekar76}. Due to
space limitations, we cannot repeat the arguments leading to that equation
here. For details about this calculation, the reader is referred 
to~\cite{Eigenetal89}. We use the same notations as are used there. Additionally, we
define the error rate $R=1-q$, which gives the probability that a
single symbol is copied erroneously. The string length will be denoted
by~$l$ throughout this paper.

The vector of the unnormalized sequence concentrations $\mb y(t)$
evolves according to
\begin{equation}\label{eq:discrete-quasispecies}
  \mb y(t+\Delta t)=\Big[\Delta t\, \mat W(t) + \Idmat\Big] \mb y(t).
\end{equation}
Here, $\mat W(t)$ is the replication matrix $\mat W(t)=\mat Q\mat
A(t)-\mat D(t)$. We assume the matrix $\mat W(t)$ is periodic with
period $T=n\Delta t$, $n \in \mathbb{N}$, with some chosen discretization time step
$\Delta t\ll T$.
After iteration of Eq.~(\ref{eq:discrete-quasispecies}), we obtain for
$t'=t+\zeta\Delta t$, $\zeta \in\mathbb{N}$
\begin{equation}\label{eq:discrete-iteration}
  \mb y(t')=\mathcal{T}\left\{\prod_{\nu=0}^{\zeta-1}\Big[\Delta t\, \mat
  W\left(t+\nu\Delta t\right) + \Idmat\Big]\right\}\,\mb y(t).
\end{equation}
where $\mathcal{T}\{.\}$ stresses that the product has to be evaluated
in the time order given by the iteration. With the definition of the
matrix (or operator)
\begin{equation}\label{eq:def-X}
  \mat X:=\mathcal{T}\left\{\prod_{\nu=0}^{n-1}\Big[\Delta t\, \mat
  W\left(\nu\Delta t\right) +
  \Idmat\Big]\right\}\,.
\end{equation}
which maps $\mb y(0)$ onto $\mb y(T)$, we are now able to write down
the solution of the discretized differential equation 
Eq.~(\ref{eq:discrete-quasispecies}) for the initial condition $\mb
y(0)$ as
\begin{equation}\label{eq:discrete-complete-sol}
  \mb y(t) =\mathcal{T}\left\{\prod_{\nu=0}^{\zeta-1}\Big[\Delta t\, \mat
  W\left(\nu\Delta t\right) +
  \Idmat\Big]\right\} \mat X^m \mb y(0)\,.
\end{equation}
where the time $t$ has been subdivided into $t=mT+\zeta\Delta t$, with
$\zeta<n$ and $m, \zeta \in\mathbb{N}$.

If we observe the system in time steps
of the period length $T$, the system appears to evolve in a static
replication landscape, which is defined by $X$. The asymptotic steady
state for the oscillation phase $\zeta=0$ is therefore given by the normalized 
Perron eigenvector $\mb \phi_{0}$ of $X$~\cite{Perron07}. For $0<\zeta<n$, the steady
states are found by application of $\mathcal{T}\{\ldots\}$ from
Eq. (\ref{eq:discrete-complete-sol}) to $\mb \phi_{0}$ and subsequent normalization.


Let us now study quantitatively the effects a periodic replication
landscape has on the prominent quasispecies. As the first step into
that direction, we start
from the Swetina-Schuster landscape~\cite{SwetinaSchuster82} and
introduce small oscillations in the master sequence's replication
coefficient~$A_0$. For reasons of simplicity, we set all decay
constants equal $D_i(t)=D$, because then they drop out of Eq. 
(\ref{eq:discrete-quasispecies}) during the foregoing linearization.

We will write the time-dependent replication coefficient~$A_0(t)$ as
\begin{equation}\label{eq:basic-timedep}
  A_0(t)=A_{\rm 0,S}\exp[\epsilon f(t)]\,,
\end{equation}
where $A_{\rm 0,S}$ is the replication coefficient in the static
landscape, $f(t)$ is a $T$-periodic function and $\epsilon$ is the
oscillation amplitude.
For $\epsilon=0$ the corresponding static landscape is reached.
The other replication coefficients are equal $A_1=\dots=A_l=A$ and
constant. We will choose $A$ 
so small that the condition $A\ll A_0(t)$ is satisfied for all~$t$ 
and $\epsilon\ll 1$. This
assures that we see a clear transition from the static case to
the dynamic case, and additionally, that the changes in the master
sequence's abundance can be directly related to the changes in~$A_0(t)$.

One of the simplest forms the function $f(t)$ in
\eqref{eq:basic-timedep} can take on is
\begin{equation}\label{eq:sin-timdep}
  f(t)=\sin(\omega t)\qquad\mbox{with}\quad\omega=2\pi/T.
\end{equation}
In the following, we will shortly discuss the influences
of different frequencies $\omega$ and amplitudes $\epsilon$ for
this time dependency.

\begin{figure}[tb]
\centerline{
 \epsfig{file={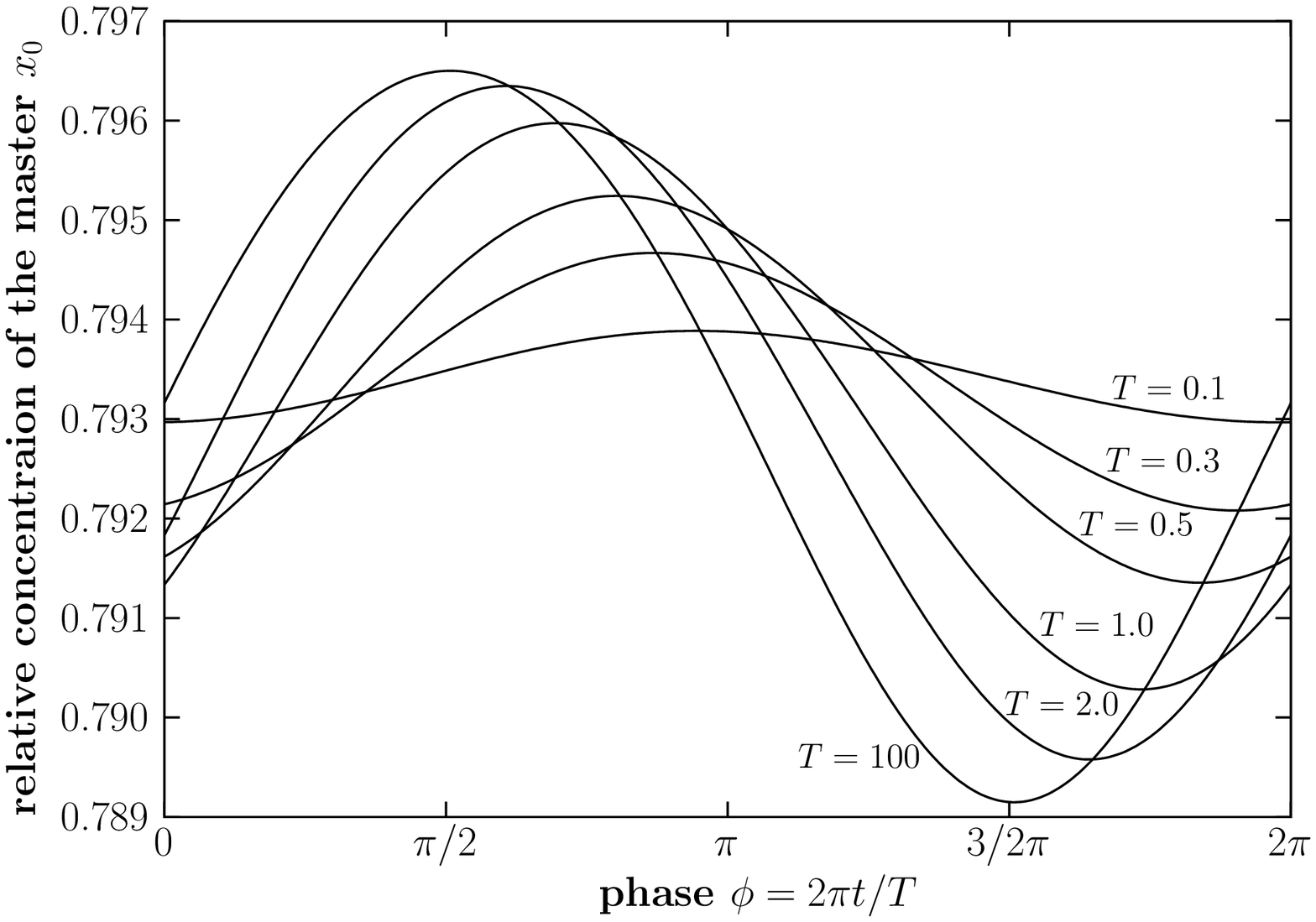}, width=8cm}
}
\caption{\label{fig:l2R0.1}The steady state oscillations of the master
 sequence in cyclically changing environments with different
 oscillation periods~$T$. Parameters used are $l=2$, $A_{\rm
 0,S}=e^{2.4}$, $A_1(t)=A_2(t)=1$, $\epsilon=0.2$.
}
\end{figure}

As response to the oscillation of the replication coefficients, a modified
oscillation is found in the concentration $x_{0}$ of the Master
sequence (Fig~\ref{fig:l2R0.1}). 
For increasing frequency $\omega$, the amplitude of the $x_{0}$
oscillation decreases and a phase shift strengthens. This behaviour is
due to the finite time a reaction system as described by Eigen's equation
needs to settle into equilibrium. In constant environments, the
asymptotic species distribution is approached in exponential time, with the relaxation
time scale $\tau$ set by the difference between the largest and the
second-largest eigenvalue of $W$. For the oscillating environments
the relaxation time needs to be compared to the period $T$. If
$T\gg\tau$, the system is virtually in equilibrium for arbitrary
(asymptotic) times $t$, whereas for $T\approx\tau$ the changes cannot
be tracked anymore and phase shift as well as amplitude damping of
the response sets in. For $T\ll\tau$ the response amplitude gets fully
damped and the system gets identical to one with the time-averaged
replication coefficients. Interpreting the $A_{0}$ and $x_{0}$ time-dependence as
input and output signal, the system acts as a low pass analog filter,
in analogy to observations made in population genetics models with
dynamic fitness landscapes \cite{SasakiIwasa87,IshiiMatsudaIwasaSasaki89,Hirst97a,HirstRowe99}. Moreover, for small $\epsilon$, the filter works linear,
which means that a sinusoidal oscillation is found in $x_{0}$, whereas
this linearity is quickly destroyed for increasing
$\epsilon$\footnote{Details on this analysis can be found 
in~\cite{Wilke99}.}.

\begin{figure}[t]
\centerline{
\input{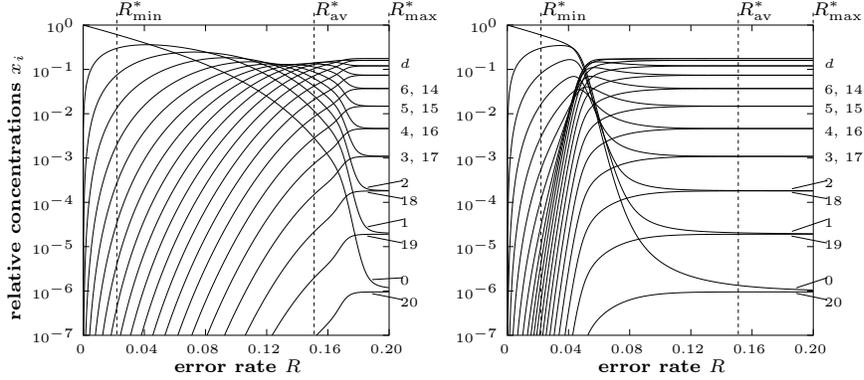}
}
\caption{\label{fig:oscet}The quasispecies distribution as a
function of the error rate $R$. $A_{0,S}=e^{2.4}, A=1, \epsilon=2, T=100$.
Two different oscillation phases are shown. {\it left\/}:
$\zeta=n/2$,
{\it right\/}: $\zeta=0$.}
\end{figure}

We will now focus on the influence of a time-dependency as given in 
Eq.~(\ref{eq:basic-timedep}) onto the error-threshold.
In accordance with the above, we have to distinguish between different
dynamic regimes. For
$T\ll\tau$, a sharp error-transition occurs at $R^{*}_{\rm av}$, which
denotes the error-threshold of a system with time-averaged
$\tilde{A}_{0}=\frac{1}{T}\smash{\int_{0}^{T}}A_{0}(t)\,dt$. Contrasting to
that, a moving error-transition can be found for $T\gg\tau$ at
approximately $R^{*}(t)$,  which denotes for any given $t$ the
error-threshold in  a constant landscape with
$\tilde{A}_{0}=A_{0}(t)$. $R^{*}(t)$ lies
between $R^{*}_{\rm max}$ and $R^{*}_{\rm min}$, which correspond to 
$\tilde{A}_{0}=\max_{t}A_{0}(t)$ and
$\tilde{A}_{0}=\min_{t}A_{0}(t)$, respectively.  In the intermediate cases
$T\approx\tau$, the numerical simulations (see Fig.~\ref{fig:oscet}) show
that the error-threshold
$R^{*}(t)$ oscillates within a smaller interval than
$[R^{*}_{\min},R^{*}_{\max}]$. 
 
\begin{figure}[b]
\centerline{
\epsfig{file={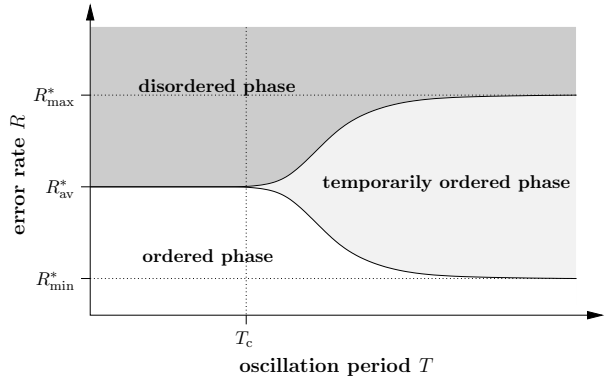}, width=8cm}
}
\caption{\label{fig:phases}Schematic phase diagram for a
  time-dependency like Eq.~(\ref{eq:basic-timedep}).}
\end{figure}

These findings allow to draw a phase diagram as displayed in
Fig.~\ref{fig:phases}. For low~$T$, we observe the standard separation
into an ordered phase (below the error threshold) and a disordered
phase (above the error threshold). With increasing oscillation period
$T$, a third, new phase appears between the two. In this phase, we
observe---for a fixed error rate $R$---an alternation between a fully
developed quasispecies and a completely disordered system. The
population seems to be moving back and forth over the error
threshold. Therefore, we call this new phase the temporarily ordered
phase. Since a similar phase diagram can be expected for any periodic
landscape with finite $R^{*}_{\rm av}$ and $R^{*}_{\rm min}\neq
R^{*}_{\rm max}$, we believe that such observations could also be made
in typical AL simulations such as Tierra or Avida~\cite{Adami98}
provided with the appropriate replication landscape. The temporarily
ordered phase would for a finite population in a rugged landscape have
the effect of causing a random drift over the landscape at some times
and a localization around a local master sequence at other times. 

Upon
completing this work, we became aware of Ref~\cite{NilssonSnoad99},
in which a different approach towards dynamic replication landscapes
is given, using a stochastic time dependency in the landscape. The results
presented there cannot directly be related to our findings here, because the
equivalent to $R^\ast_{\rm av}$ vanishes in~\cite{NilssonSnoad99},
while the equivalents to $R^\ast_{\max}$ and $R^\ast_{\min}$ take on
the same finite value. This leads to a different phase
diagram than the one we observe here.

\begin{thebibliography}{10}

\bibitem{Eigen71}
M. Eigen, Naturwissenschaften {\bf 58},  465  (1971).

\bibitem{EigenSchuster79}
M. Eigen and P. Schuster, {\em The Hypercycle---A Principle of Natural
  Self-Organization} (Springer-Verlag, Berlin, 1979).

\bibitem{Eigenetal88}
M. Eigen, J. McCaskill, and P. Schuster, {\frenchspacing J. Phys. Chem.} {\bf
  92},  6881  (1988).

\bibitem{Eigenetal89}
M. Eigen, J. McCaskill, and P. Schuster, {\frenchspacing Adv. Chem. Phys.} {\bf
  75},  149  (1989).

\bibitem{Biebricheretal83}
C.~K. {Biebricher}, M. Eigen, and W.~C. {Gardiner, Jr.}, Biochemistry {\bf 22},
   2544  (1983).

\bibitem{Jones79a}
B.~L. Jones, {\frenchspacing Bull. Math. Biol.} {\bf 41},  761  (1979).

\bibitem{Jones79b}
B.~L. Jones, {\frenchspacing Bull. Math. Biol.} {\bf 41},  849  (1979).

\bibitem{Demetriusetal85}
L. Demetrius, P. Schuster, and K. Sigmund, {\frenchspacing Bull. Math. Biol.}
  {\bf 47},  239  (1985).

\bibitem{ThompsonMcBride74}
C.~J. Thompson and J.~L. McBride, {\frenchspacing Math. Biosci.} {\bf 21},  127
   (1974).

\bibitem{JonesEnnsRangnekar76}
B.~L. Jones, R.~H. Enns, and S.~S. Rangnekar, {\frenchspacing Bull. Math.
  Biol.} {\bf 38},  15  (1976).

\bibitem{Perron07}
O. Perron, {\frenchspacing Math. Ann.} {\bf 64},  248  (1907).

\bibitem{SwetinaSchuster82}
J. Swetina and P. Schuster, {\frenchspacing Biophys. Chem.} {\bf 16},  329
  (1982).

\bibitem{SasakiIwasa87}
A. Sasaki and Y. Iwasa, Genetics {\bf 115},  377  (1987).

\bibitem{IshiiMatsudaIwasaSasaki89}
K. Ishii, H. Matsuda, Y. Iwasa, and A. Sasaki, Genetics {\bf 121},  163
  (1989).

\bibitem{Hirst97a}
T. Hirst,  in {\em Fourth European Conference on Artificial Life}, edited by P.
  Husband and I. Harvey (MIT Press, Cambridge, MA, 1997), pp.\ 425--431.

\bibitem{HirstRowe99}
A.~J. Hirst and J.~E. Rowe, {\frenchspacing J. theor. Biol.}  (1998),
  submitted.

\bibitem{Wilke99}
C.~O. Wilke, Evolutionary Dynamics in Time-Dependent Environments (Ph.D. thesis, Ruhr-Universit\"at Bochum, 1999).
  http://www.neuroinformatik.ruhr-uni-bochum.de/ini/PEOPLE/wilke/ps/PhD.ps.gz.

\bibitem{Adami98}
C. Adami, {\em Introduction to Artificial Life} (Telos, Springer-Verlag
  Publishers, Santa Clara, 1998).

\bibitem{NilssonSnoad99}
M. Nilsson and N. Snoad,
  eprint physics/9904023 (1999).

\end{thebibliography}

\end{document}